\documentclass[twocolumn,showpacs,preprintnumbers,amsmath,amssymb,pre]{revtex4}

\usepackage[latin1]{inputenc}
\usepackage[T1]{fontenc}
\usepackage{ae,aecompl}

\usepackage{amsmath}
\usepackage[fleqn,tbtags]{mathtools}

\usepackage{graphicx}

\newcommand{\beq}{\begin{eqnarray}}
\newcommand{\eeq}{\end{eqnarray}}
\newcommand{\Req}[1]{Eq. (\ref{E#1})}
\newcommand{\req}[1]{(\ref{E#1})}

\newcommand{\Rfg}[1]{Fig. \ref{F#1}}

\newcommand{\Rtb}[1]{Table \ref{T#1}}

\begin{document}

\title{Weak Nanoscale Chaos And Anomalous Relaxation in DNA}

\author{Alexey K. Mazur}
\affiliation{UPR9080 CNRS, Universit\'e Paris Diderot, Sorbonne Paris Cit\'e,\\
Institut de Biologie Physico-Chimique,\\
13, rue Pierre et Marie Curie, Paris, 75005, France}


\begin{abstract}
Anomalous non-exponential relaxation in hydrated biomolecules is
commonly attributed to the complexity of the free-energy landscapes,
similarly to polymers and glasses.
It was found recently that the hydrogen-bond breathing of terminal DNA
base pairs exhibits a slow power-law relaxation attributable to weak
Hamiltonian chaos, with parameters similar to experimental data.
Here, the relationship is studied between this motion and spectroscopic
signals measured in DNA with a small molecular photoprobe inserted into the
base-pair stack.
To this end, the earlier computational approach in combination with
an analytical theory is applied to the experimental DNA fragment.
It is found that the intensity of breathing dynamics is strongly
increased in the internal base pairs that flank the photoprobe, with
anomalous relaxation quantitatively close to that in terminal base pairs.
A physical mechanism is proposed to explain the coupling between
the relaxation of base-pair breathing and the experimental response
signal.
It is concluded that the algebraic relaxation observed experimentally
is very likely a manifestation of weakly chaotic dynamics of
hydrogen-bond breathing in the base pairs stacked to the photoprobe,
and that the weak nanoscale chaos can represent an ubiquitous hidden
source of non-exponential relaxation in ultrafast spectroscopy.
\end{abstract}

\maketitle

In physics, slow non-exponential relaxation is considered as a
hallmark of complexity of condensed disordered systems, notably,
polymers, glasses, and supercooled liquids.
Such kinetics is called anomalous and it results from dynamics on
complex energy landscapes with many coupled degrees of freedom
\cite{Klafter:86}.
The same interpretation is commonly applied to anomalous relaxation in
hydrated proteins \cite{Iben:89,Frauenfelder:91,Fenimore:05} and
double helical DNA \cite{Brauns:02,Andreatta:05}.
We found recently that the hydrogen-bond (HB) breathing of terminal
DNA base pairs exhibits a slow power-law relaxation attributable to
weak Hamiltonian chaos \cite{Mzprl:15}.
Its parameters appeared similar to the experimental data
obtained by time-resolved fluorescence Stokes shift (TRFSS) spectroscopy
\cite{Andreatta:05}.
Since the fluorescent probes are never placed near DNA ends the
specific motion we studied could not play a role in earlier
experiments and, strictly speaking, the above agreement was
accidental.
Nevertheless, the standard qualitative interpretation of these
experimental data is very controversial and there are reasons to
believe that localized weakly chaotic modes represent the true
physical cause of anomalous relaxation in hydrated biomolecules where
this phenomenon was revealed by small molecular probes
\cite{Mzprl:15}.

Power law distributions are ubiquitous in physics because this law is
a large number limit for sums of random variables with infinite
variances \cite{Shlesinger:93,Klafter:96,Eliazar:12}, that is, it
plays a universal role similar to that of the Gaussian.
Therefore, when two processes share a power-law form of time
autocorrelation functions, it just means that both processes involve
substates with broad lifetime distributions, but not necessarily
coupled.
The similarity of exponents is a stronger evidence, but these values are
estimated with limited accuracy, whereas they are not very different
between many distinct processes.
For a casual relationship between the base-pair breathing
and the TRFSS data this motion should occur in a close
vicinity of the photoprobe and this dynamics should modulate a strong
applied electrostatic field.
Both these conditions are problematic.
The opening of internal DNA base pair should be negligible under
relatively low temperatures used in TRFSS experiments (15$^\circ$C,
\cite{Brauns:02}).
Besides, dynamics of internal base pairs is strongly restricted and
its parameters can differ from those we obtained for terminal bases
\cite{Mzprl:15}.
The origin of the strong electrostatic field modulated by base-pair
breathing is also unclear.
To shed light upon the above problems here the base-pair
breathing is studied in an experimental DNA fragment by
combining the earlier computational approach \cite{Mzprl:15}
with a new analytical theory.
The intensity and the slow relaxation of these dynamics are
studied in the internal base pairs that flank the coumarin
photoprobe.
An alternative mechanism of coupling between these motions and
the TRFSS response signal is proposed.

\begin{figure}[ht]
 \centerline{\includegraphics[width=0.48\textwidth]{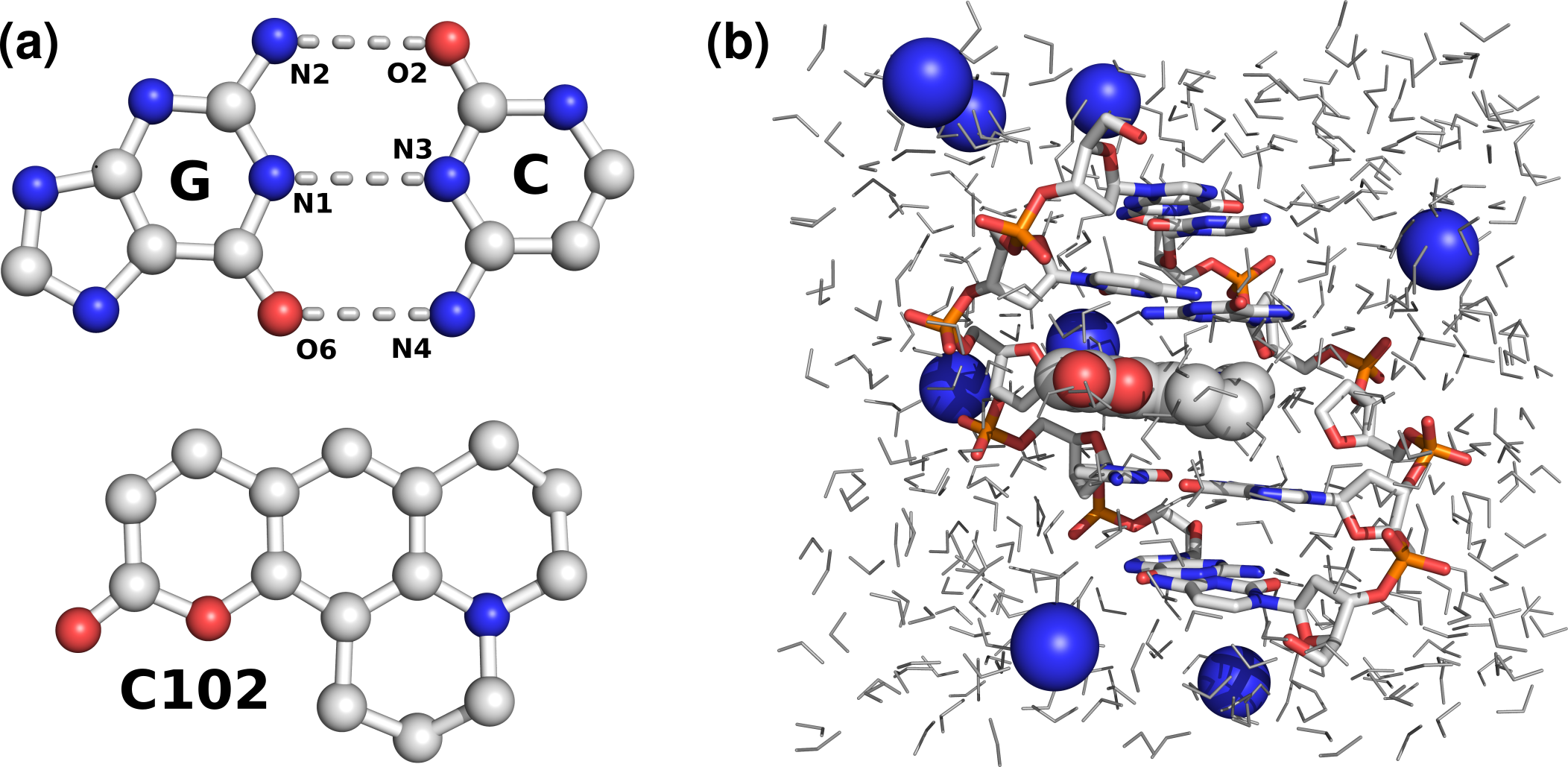}}
\vglue -0.3cm
\caption{\label{Fmol}(Color online)
{\bf (a)} The stack components: a GC base pair, with three H-bonds
indicated by thick dashed lines (top), and a coumarin C102 residue
(bottom), both shown with the major groove edges facing down.
{\bf (b)} The model system used in MD simulations.  The DNA structure
is shown by sticks and the hydration water is displayed by thin lines.
The coumarin residue and the sodium ions are highlighted by spheres.
The standard DNA atom color coding is used, that is, gray, red, blue,
and orange for carbons, oxygens, nitrogens, and phosphates,
respectively.  }
\end{figure}

The model system is designed as shown in \Rfg{mol}.  A DNA duplex is
placed in a periodic box with 456 water molecules and eight sodium
ions. The sequence of this DNA is $^{\texttt{\normalsize
GCMCG}}_{\texttt{\normalsize CG\_GC}}$ where M stands for coumarin.
In \Rfg{mol}b the duplex is shown from the major groove side; the
opposite groove is called minor. This molecule is a representative
fragment of the experimental DNA \cite{Brauns:02,Andreatta:05}. The
photoprobe (coumarin C102) is inserted into the middle of the stack
formed by four base-pairs, with two oxygens exposed into the major
groove.  This photoprobe has approximately the size of a base pair and
it is not exactly planar. The DNA end fraying was prevented by
flat-bottom restraints applied to hydrogen bonds in the terminal base
pairs.  All-atom MD simulations were carried out as described earlier
\cite{Mzprl:15}.  Other details can be found in Appendix A.

The HB-breathing was followed by measuring the statistics of
Poincar\'e recurrences for distances ($R$) between atoms that form
Watson-Crick (WC) H-bonds. The celebrated Poincar\'e recurrence
theorem of 1890 \cite{Poincare:1890c} guarantees that a dynamic
trajectory with a fixed energy and bounded phase space will always
return in a close vicinity of the initial state. The statistics of
recurrences is defined as the probability distribution $P(\tau)$ of
returns with times longer than $\tau$.
Hamiltonian dynamics can be regular and chaotic
\cite{Lichtenberg:92}. In purely chaotic phase spaces any
trajectory rapidly fills accessible areas and $P(\tau)$ drops
exponentially, like the probability of any region to remain
unvisited during time $\tau$ \cite{Arnold:68,Cornfeld:82}.
In contrast, in mixed phase spaces where islands of stable regular
motion are embedded in a chaotic sea $P(\tau)$ decays as
\cite{Chirikov:81,Chirikov:84}
\beq\label{Epow}
P(\tau) \propto 1/\tau^{\beta} \;\; .
\eeq
This occurs because the stability islands are commonly covered by boundary
layers of fractal structures formed by smaller islands, and chaotic
trajectories are trapped in these layers for very long periods
\cite{Zaslavsky:99}. This complexity is intrinsic in Hamiltonian
systems of any size. The power-law decay is persistent in canonical
models with one or several degrees of freedom
\cite{Ding:90,Fendrik:95,Altmann:07,Shepelyansky:10,Altmann:13}.
For canonical model systems, the Poincar\'e exponent is
$\beta \sim 1.5$
\cite{Chirikov:81,Karney:83,Chirikov:84,Meiss:85,Chirikov:99,Cristadoro:08},
but in real dynamics it may vary with the interaction potential
\cite{Mzprl:15}.
Probably the simplest mechanical example is the three-body
Coulomb problem \cite{Schlagheck:01}. In large heterogeneous
systems like hydrated macromolecules the algebraic decay is
overshadowed by numerous chaotic modes, but it can manifest
itself in special conditions \cite{Mzprl:15}.

In DNA dynamics, the stopwatch is started when a given distance
exceeded a certain threshold ($R_{th}$) and stopped once the boundary
is crossed in the opposite direction.  $P(\tau)$ is obtained by
counting the number of recurrences with duration larger than $\tau$
and normalizing it by the total number of events, that is, $P(0)=1$ by
construction.  The fraction of time spent in the opened state is
evaluated simultaneously. These computations are trivially
parallelizable, that is, the $P(\tau)$ statistics can be accumulated
in a large number of independent MD trajectories. The results
discussed below were obtained by using parallel computations on 129
cores for a model system with 3557 degrees of freedom for 1665
atoms to obtain the total sampling of about 100 $\mu s$.

\begin{table}[ht]\caption{\label{Tpop1}
Populations (in ppm) of opened HB-states for different GC base pairs.
The internal and terminal base pairs in unmodified tetramer DNA
\cite{Mzprl:15} are coded as I and T, respectively. The two pairs
stacked to the photoprobe are denoted according to the DNA strand
direction, that is, 5'n (5'-neighbor, upper in \Rfg{mol}b) and 3'n
(lower), respectively.
}
\centerline{\begin{tabular}[t]{|c|c|c|c|c|}  \hline
 HB    &  I   &  T   &  5'n  &  3'n  \\ \hline
O6-N4  &  9   &1043  &  22   & 7868  \\
N1-N3  &  0   & 197  &   7   & 5033  \\
N2-O2  &  1   &  46  &   3   &12005  \\ \hline
\end{tabular}}
\end{table}

\Rtb{pop1} contains populations of partially opened states of GC pairs
in different locations. These populations were measured in MD
as fractions of trajectory time spent with $R>R_{th}$=4.15 \AA\ for
the three H-bonds indicated in \Rfg{mol}a. The probability of opening
strongly depends upon the H-bond as well as the environment of the
base pair. In unmodified DNA the internal base pairs are very stable
compared to the terminal ones; the O6-N4 H-bonds are opened easier
than other and the breathing predominantly occurs towards the major
groove, that is, towards the viewer in \Rfg{mol}b. The HB-breathing in
internal pairs is strongly increased when they are stacked to the
intercalated coumarin. From the 5'-side of the photoprobe this
increase is moderate, but the opposite 3'n pair is perturbed very
strongly. The HB-breathing in this GC pair qualitatively differs from
others because the N2-O2 and O6-N4 H-bonds appear more labile than the
central one, which means that partial openings occur towards both
grooves.

\begin{figure}[ht]
 \centerline{\includegraphics[width=0.48\textwidth]{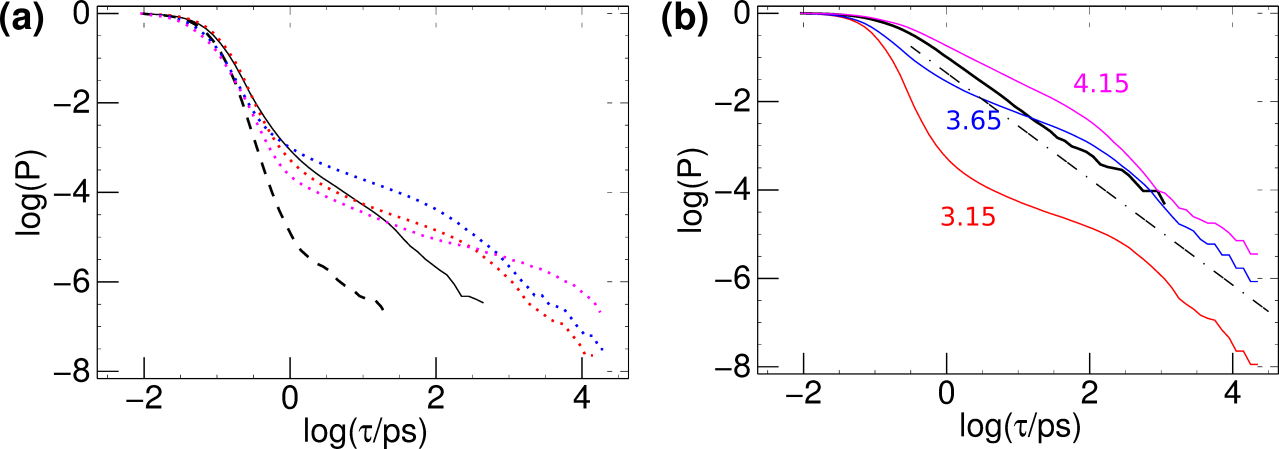}}
\vglue -0.3cm
\caption{\label{Frec1}(Color online)
{\bf (a)} The statistics of Poincar\'e recurrences $P(\tau)$ for five
WC H-bonds ($R_{th}= 3.15$ \AA). The black solid and dashed lines
correspond to O6-N4 bonds in terminal and internal base pairs,
respectively, in unmodified DNA \cite{Mzprl:15}. The colored dotted
traces correspond to the three H-bonds of the 3'n pair in \Rtb{pop1}.
{\bf (b)} $P(\tau)$ profiles corresponding to the O6-N4 bond of the
3'n pair computed for three different thresholds $R_{th}$ indicated in
the figure. The solid black line shows the corresponding trace for
terminal base pairs in unmodified DNA ($R_{th}= 4.15$ \AA).  The
dash-dotted straight line shows the power-law decay with the exponent
$\beta=1.2$.  The logarithms are decimal.
}\end{figure}

\Rfg{rec1}a displays the statistics of recurrences to threshold
$R_{th}= 3.15$ \AA\ for a few representative H-bonds.  For internal
base pairs the $P(\tau)$ profile is nearly exponential until
$\tau\approx1$ ps. The exponential regime is similar for all H-bonds.
It results from rapid chaotic oscillations within bonded ground states
because the saddle point of the H-bond potential occurs well beyond
3.15 \AA. Slower returns to $R_{th}$ are due to trajectories that
wander outside the ground state valley and they give an algebraic
decay of $P(\tau)$. As shown in \Rfg{rec1}b the relative weight of
slow returns naturally grows with $R_{th}$. An example of a partially
opened structure is shown in \Rfg{toc}. It is seen that in the upper
base pair the hydrogen bond at the major groove side is broken by a
stable bridging water molecule in direct contact with coumarin. Such
structures become possible when the hydrogen bond is stretched by more
than one angstrom, i.e. to the length of about 4 \AA. As seen in
\Rfg{rec1}b this distance is the threshold beyond which the
exponential relaxation becomes negligible.

\begin{figure}[ht]
 \centerline{\includegraphics[width=0.3\textwidth]{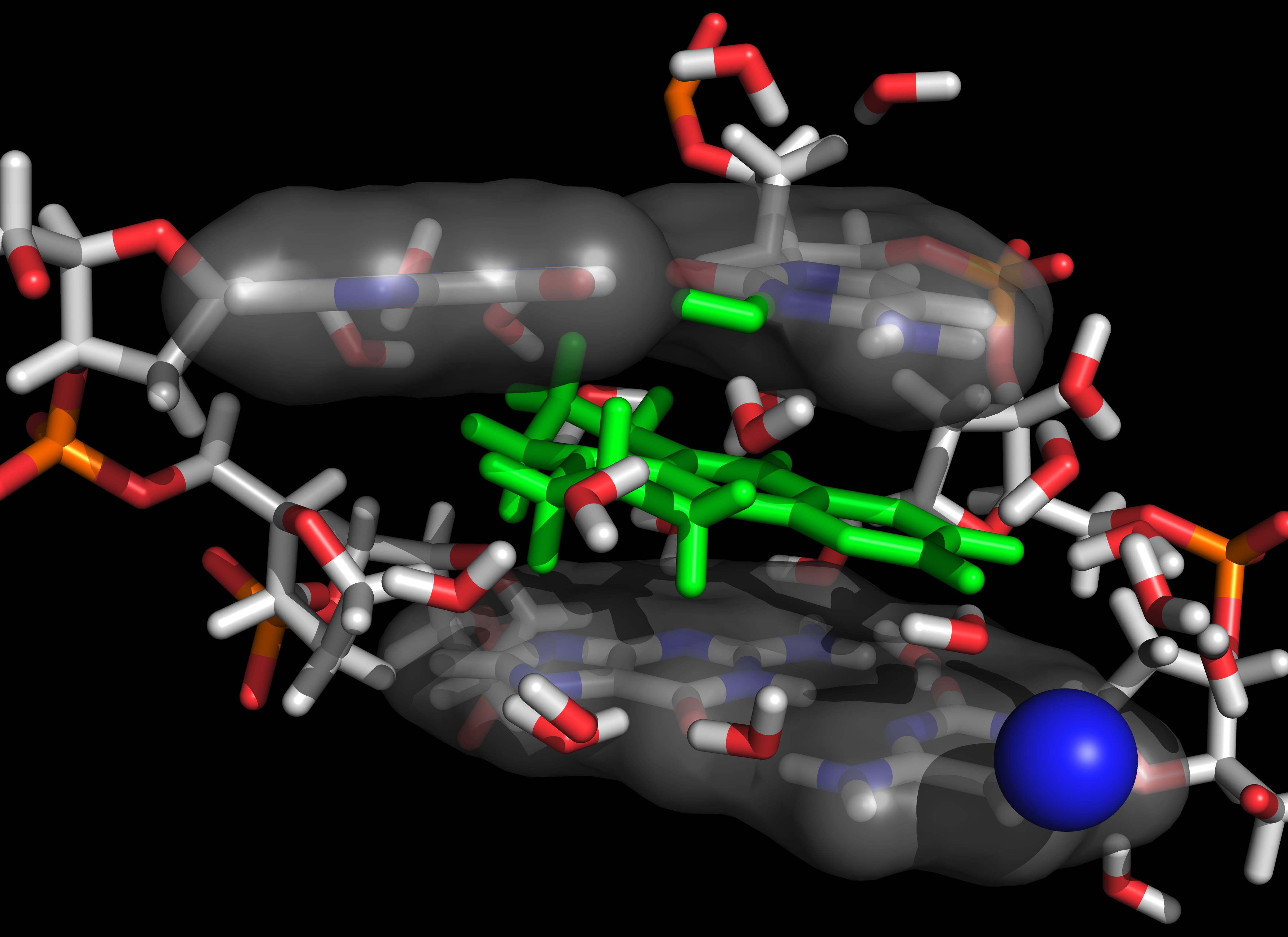}}
\vglue -0.3cm
\caption{\label{Ftoc}(Color online)
A snapshot from MD simulation of the model DNA fragment with a
partially opened base pair stacked to coumarin. The coumarin residue
is highlighted by green.  It is seen that the O6-N4 hydrogen bond is
broken and replaced by a bridging water molecule also highlighted by
green.  Other water molecules shown and the sodium ion (blue sphere)
take part in the coumarin's hydration shell in the major groove.
}\end{figure}

The transition between the exponential and algebraic regimes depends
upon the H-bond and the base pair location, but eventually the
power-law decay \Req{pow} is established with an exponent close to
$\beta=1.2$ obtained for end fraying \cite{Mzprl:15}. This is shown in
\Rfg{rec1} and \Rfg{rec2}a, respectively, for 3'n and 5'n pairs from
\Rtb{pop1}.  At the same time, the HB-breathing in the coumarin's
neighbors cannot be similar to the fraying of terminal base pairs, and
it evidently differs, which is seen in the shapes of the plots in
\Rfg{rec1} and the relative populations of open states in \Rtb{pop1}.
In addition, \Rfg{rec2}b shows that the end fraying is accompanied by
a faster growth of the average distance between the reference atoms
with the life time of the opened state.  These results argue against
the mechanisms of the fractional Brownian motion \cite{Voss:92} in the
dynamics of HB-distances and the corresponding stochastic origin of
the power-law decay. Such models would produce the growth $\langle
R\rangle\propto\tau^H$, with the Hurst index $H$ defining the exponent
$\beta$ in \Req{pow}. In contrast, \Rfg{rec2}b reveals that this
growth is nearly linear in semi-logarithmic coordinates. Moreover, the
value of $\beta$ is robust against the differences in the growth
rates, which agrees with the suggestion based upon the chaos theory
that $\beta$ is determined by the degree of the polynomial that
approximates the energy profile near the saddle points
\cite{Mzprl:15}. This parameter should not depend upon the base pair
location.

\begin{figure}[ht]
 \centerline{\includegraphics[width=0.48\textwidth]{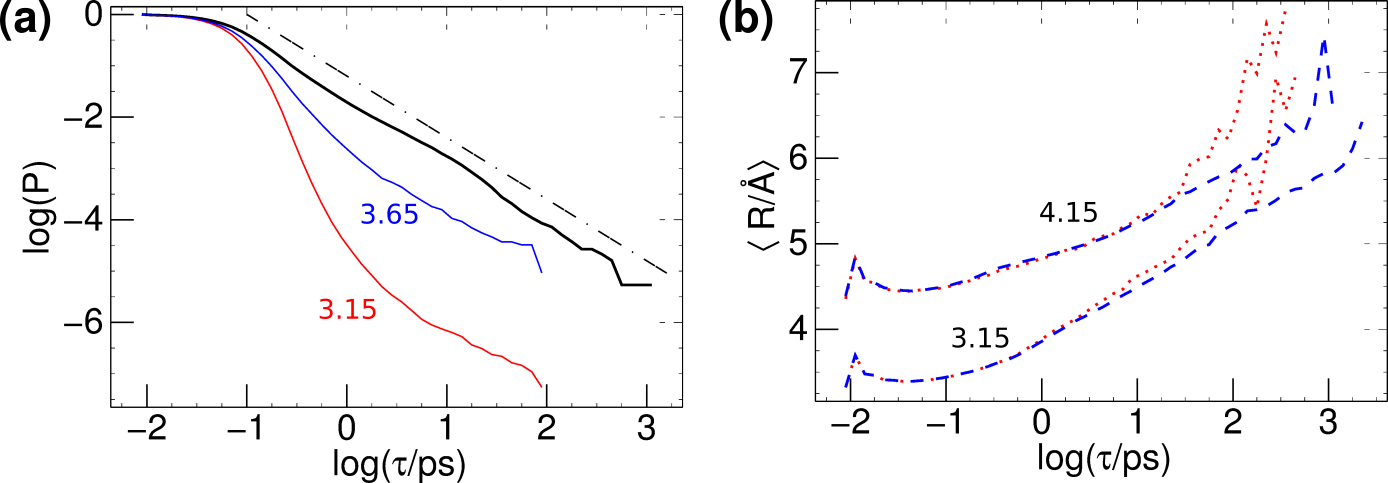}}
\vglue -0.3cm
\caption{\label{Frec2}(Color online)
{\bf (a)} $P(\tau)$ profiles corresponding to the O6-N4 bond of the
5'n pair in \Rtb{pop1} computed for two thresholds $R_{th}$ indicated
in the figure. The solid black line shows a similar trace for terminal
base pairs in unmodified DNA ($R_{th}= 3.55$ \AA). The dash-dotted
straight line shows the power-law decay with the exponent $\beta=1.2$.
{\bf (b)} The dependencies of average distances $\langle R\rangle$ for
O6-N4 H-bonds in the 3'n pair (dashed blue) and terminal base pairs
(dotted red) for two thresholds indicated in the figure. The
logarithms are decimal.
}\end{figure}

The power-law relaxation in DNA has been revealed by TRFSS in a broad
time range starting from sub-picoseconds \cite{Brauns:02,Andreatta:05}.
This phenomenon was analyzed with different theories
\cite{Kalosakas:06,Sen:09,Pal:10,Furse:11}, but its origin remains
controversial.  The DNA double helix has only a few well-studied
conformational substates with spatially localized dynamics that poorly
fits the energy landscape description.  There are slow modes in the
conformational dynamics of coumarin and DNA, but the possible
candidates \cite{Furse:11} involve only one or a few local minima and
they cannot result in broad lifetime distributions.  In contrast, the
hydration dynamics can be complex, but it does not have relaxation
modes beyond the picosecond range \cite{Halle:09,Furse:10a}.  At
present, there is no agreement even on whether this effect is due to
DNA itself, or the hydration water, or both
\cite{Berg:08,Halle:09,Furse:10a}.

The optical $S0\rightarrow S1$ transition in coumarins instantaneously
increases the dye's dipole moment and perturbs the electrostatic
equilibrium of its environment \cite{Bagchi:10}.  The Stokes effect
occurs due to re-equilibration the environment during the excitation
lifetime and results in a fluorescence red shift.  The four bases
stacked to the photoprobe in DNA represent the major part of its
environment, therefore, the HB-breathing and the fluorescence
transitions are likely to be coupled. However, a causal relationship
between the HB-breathing and the power-law relaxation revealed in
experiments is not at all evident. The measured signal is
\beq\label{ESt}
S(t)=\frac{\nu(t)-\nu(\infty)}{\nu(0)-\nu(\infty)}
\eeq
where $\nu(t)$ is the fluorescence emission frequency and $t$ is the
time after laser excitation. For theoretical analysis $S(t)$ is
commonly approximated as
\beq\label{ES=C}
S(t)\approx C(t)=\frac{\langle\delta E(0)\delta E(t)\rangle}
 {\langle\delta E(0)^2\rangle}
\eeq
where $E(t)$ and $C(t)$ represent the solvation energy and its
equilibrium autocorrelation function, respectively. This splits the
experimental response into a sum of autocorrelation functions of independent
electrostatic contributions. To account for the experimental data this
sum must involve a power-law term with an exponent around 0.15
\cite{Andreatta:05}. A similar value is obtained from \Req{pow} with
$\beta=1.2$ by using an estimate proposed in the chaos theory
\cite{Karney:83,Chirikov:99}
\beq\label{ECtau1}
C(\tau) \propto \tau P(\tau),
\eeq
but this is just a curious mathematical coincidence (see Appendix B).
Indeed, the power spectral density of time fluctuations for
HB-distances considered here decays with an exponent
$\eta=1.63$ \cite{Mzprl:15}, which means that in \Req{S=C} the
effect of a hypothetical electrostatic field modulated by
the base-pair breathing would disagree with the available
experimental data.

The physical coupling between the measured TRFSS signal and the
base-pair breathing can have a completely different origin.
This alternative mechanism is described below.
Although the effect is non-linear and complex its physics is similar to
that in a simple two-state system described by linear kinetics.
In the most relevant case, when the lifetime of the opened base-pair
state is very small, a general solution is also available.
The following derivations employ a perturbative approach in the sense
that the qualitative similarity with linear system is used as long as
possible, with the main result derived by applying the same ideas in
the principal limiting case.
The general solution for the linear system and the underlying
assumptions are discussed in larger detail in Appendix C.

A photoprobe can receive a photon when the neighboring base pairs are
closed ($A$) or partially opened ($B$). The excited coumarins are
partitioned between the two states with concentrations $C_\textsc{A}$ and
$C_\textsc{B}$, respectively. In the linear approximation the balance
between them is described by schema
\beq\label{EAB}
0\xleftarrow[]{k_0}
A\xrightleftharpoons[k_{-}]{k_{+}}B\xrightarrow[]{k_0}0,
\eeq
where $k_{+}$ and $k_{-}$ are the rate constants of base pair opening
and closing, respectively, and the quenching constant $k_0$ determines
the excitation lifetime. In states $A$ and $B$ the solvent relaxation
and the red-shifted emission occur according to the conventional
scenario described by \Req{St} and \req{S=C}, however, the fluorescence
parameters in these two states can differ. The excitation and
emission spectra of coumarin in water are significantly shifted from
those in DNA, and the excitation lifetime is much shorter
\cite{Brauns:99}. Similar changes are observed when DNA samples with
intercalated coumarins are melted \cite{Brauns:99}. In state $B$ new
water molecules like that in \Rfg{toc} come in between the bases and,
due to their high polarity and direct contact with coumarin, they can
alter the fluorescence parameters.

Let $f=C_\textsc{B}/(C_\textsc{A}+C_\textsc{B})$ denote the relative
population of opened pairs stacked to excited coumarins. The observed
emission frequency $\nu(t)$
in \Req{St} is
\beq\label{EnuAB}
\nu=\left(1-f\right)\nu_\textsc{A}+f\nu_\textsc{B}=
     \nu_\textsc{A}+f(\nu_\textsc{B}-\nu_\textsc{A}).
\eeq
where $\nu_\textsc{A}$ and $\nu_\textsc{B}$ are the emission frequencies
of the two states, respectively. Because the static spectral shift
$\nu_\textsc{B}-\nu_\textsc{A}$ is large compared to the maximal
amplitude of the time-dependent response \cite{Brauns:99}, even a small
time variation of $f(t)$ can be noticeable. Assuming that the solvent
relaxation is fast and $\nu_\textsc{A}$ and $\nu_\textsc{B}$ always
remain at steady state levels, \Req{St} gives
\beq\label{ESt1}
S(t)=\frac{f(t)-f(\infty)}{f(0)-f(\infty)}.
\eeq
In the linear case, the relaxation to equilibrium is exponential
\beq\label{Eft1}
f(t)=f(\infty)+\delta\exp\left(-\lambda t\right),
\eeq
\beq\label{ESt2}
\ S(t)=\exp\left(-\lambda t\right)
\eeq
where $\lambda=k_{+}+k_{-}$ is the only non-zero eigenvalue and
$\delta$ is the initial perturbation (see Appendix C).
The $S0\rightarrow S1$ excitation due to the laser pulse changes the
eigenvalues and creates a population of base pairs stacked to excited
coumarins with a perturbed $A\rightleftharpoons B$ equilibrium, which
causes a TRFSS response.

Now consider a non-linear case when the $B\rightarrow A$ transition
gives a non-exponential statistics of Poincar\'e recurrences
$P(\tau)$, which means that it is not a Markov process, and the
probability of base-pair closing effectively depends upon the time
spent in the open state (trapping time). When the system is in
detailed balance equilibrium the $A\rightarrow B$ and $B\rightarrow A$
flows are mutually compensated due to steady-state trapping time
distributions established for both opened and closed pairs. With
stationary distributions, the flows are proportional to
concentrations, which makes the problem equivalent to the linear case.
The shapes of the stationary distributions are similar in base pairs
stacked to normal and optically excited coumarins. For instance, if
the coumarin excitation lifetime is smaller in state $B$ this
sub-population will disappear more rapidly, but the trapping time
distribution is not affected because the rate of quenching does not
depend upon the previous history of the opened pair.

When the laser pulse arrives at $t=0$, the initial state of the
sub-population of base pairs stacked to excited coumarins can be
decomposed into a constant part corresponding to the new equilibrium,
and a perturbation which disappears in the course of relaxation.  The
irreversible decay of this second part corresponds to the TRFSS signal
that we want to evaluate. At $t=0$ both parts have the stationary
trapping time distributions established before the excitation.
With $t>0$, the assumption of stationarity remains valid for the first
part, but for the irreversible decay of the perturbation this
assumption fails, therefore, the equivalence with the linear problem
is lost.  However, when the lifetime of state $B$ is small compared to
that that of state $A$ the kinetics of decay is entirely determined by
the starting trapping time distribution in $B$, and the analytical
solution exists in the general case.

The main idea can be understood from the linear case. The
perturbation is an antisymmetric vector in the space of concentrations
$(C_\textsc{A},C_\textsc{B})$ because it represents an excess in one state and an equal
deficit in the other with respect to the equilibrium. Suppose
$\delta>0$ in \Req{ft1}, i.e., the equilibrium is perturbed by
transferring a small number of molecules from $A$ to $B$.  Since
$k_{-}\gg k_{+}$ this does not change the $A\rightarrow B$ flow and
increases the $B\rightarrow A$ flow by the rate of decay of the excess
in $B$. With $\delta<0$ the excess is created in $A$ and the deficit
in $B$. The $A\rightarrow B$ flow is again changed negligibly, but
the opposite flow is strongly reduced and the corresponding part of
the equilibrium $A\rightarrow B$ flow is not compensated.  As a
result, the deficit in $B$ is reduced with the same rate as the
excess the previous case, and the relaxation always looks like a decay
of the state with the smaller lifetime.

Now consider the general case. With
$\delta>0$ at time $t=0$ we have an excess of opened pairs
with a stationary trapping time distribution. In equilibrium, base
pairs are opened with a constant rate $\upsilon$. The stationary
distribution is the probability density of pairs opened already for
time $\tau$, that is
\beq
\int_\tau^\infty p(t)\upsilon dt=\upsilon P(\tau)
\eeq
where $p(t)$ is the normalized probability density of Poincar\'e returns
\beq
\int_0^\infty p(t)dt=1 .
\eeq
Let $C_\textsc{B}(t)$ denote the number of opened pairs corresponding
to the excess $\delta>0$. We have
\beq
C^0_\textsc{B}(0)\propto\int_0^\infty P(\tau)d\tau.
\eeq
This perturbation evolves as if the opening of base pairs stopped
at $t=0$. The probability that a pair already opened for time $\tau$
will stay opened during time $t$ is $P(\tau+t)$. Therefore,
the initial perturbation $C^0_\textsc{B}(0)$ decays with time as
\beq
C_\textsc{B}(t)\propto\int_0^\infty P(\tau+t)d\tau=\int_t^\infty P(\tau)d\tau.
\eeq
Substitution into \Req{St1} yields the measured response function
\beq\label{ESt3}
S(t)\propto\int_t^\infty P(\tau)d\tau
\eeq
and \Req{Ctau1} for power-law decays.  Now consider the case
$\delta<0$. At time $t=0$ we have a deficit in the equilibrium
distribution of opened pairs which is reduced by a constant flow from
the equilibrated pool of closed pairs. The equilibrium will be
reestablished when the deficit is closed, with the stationary trapping
time distribution recovered. The time dependence is computed as
\beq
C_\textsc{B}(t)\propto \int_0^t P(\tau)d\tau=const-\int_t^\infty P(\tau)d\tau,
\eeq
and substitution into \Req{St1} again yields \Req{St3}.

The above theory suggests that the power-law relaxation experimentally
observed in DNA is indeed caused by the base-pair breathing dynamics.
The equilibrium population of partially opened pairs is low, which
means that the rate of base-pair closing is much larger than that of
opening. Therefore, the decay of fluctuations is dominated by
base-pair closing in agreement with the assumption used above.
Although our model system has many degrees of freedom the
HB-breathing is essentially a one-dimensional motion. The weak dynamic
chaos is currently the only and the most reasonable explanation of the
observed algebraic relaxation. It is difficult, however, to obtain a
formal proof of this assertion because of a limited choice of
instruments of the chaos theory applicable to large systems. Further
work in this direction is necessary.

The theory developed above also clarifies an unclear issue concerning
the role of the coumarin's dipole moment. Presentations of the TRFSS
method often start from an assertion that optical $S0\rightarrow S1$
transitions strongly increase the dye's dipole moments. This is
necessary for slow relaxation in the conventional approach because a
significant area around the dye must be involved and the long-range
electrostatic interactions are indispensable \cite{Bagchi:10}.
However, for coumarin C102 shown in \Rfg{mol}a the measured increase
of the dipole moment does not exceed 40\%, with absolute values
between two and three debyes \cite{Cave:02b}. Quantum mechanics
calculations show that the dipoles of $S0$ and $S1$ states as well as
their difference are distributed over the whole molecule rather than
localized, and MD simulations with excited coumarin reveal no
difference from the ground state (unpublished). Indeed, in an aqueous
ionic environment a distributed dipole similar to that of a single
water molecule can hardly cause noticeable long-range effects.  The
alternative mechanism proposed above resolves this controversy
because it is applicable even with zero dipole moments.

In summary, it is shown that, in contrast to traditional statistical
mechanics explanations, the power-law relaxation earlier discovered in
DNA is likely to result from chaotic dynamics of a few or even one
degree of freedom of HB-breathing in the base pairs flanking the
photoprobe.  The new theory is generally applicable to ultrafast
relaxation techniques, and it may be interesting to look for instances
of weak nanoscale chaos in other controversial cases of algebraic
relaxation in different systems.

\begin{acknowledgments} The computational resources used in this
study were supported by the "Initiative d'Excellence" program from
the French State (Grant "DYNAMO", ANR-11-LABX-0011-01)".
\end{acknowledgments}

\appendix

\section{Molecular dynamics simulations}

DNA duplexes were modeled in aqueous environment neutralized by sodium
ions \cite{Joung:08}, using a recent version of the all-atom AMBER
forcefield \cite{Cornell:95,Wang:00,Perez:07a,Zgarbova:13} with SPC/E
water \cite{Berendsen:87} in periodic boundaries. The coumarin partial
charges were obtained by the RESP method \cite{Bayly:93,Cornell:93},
with sugar and phosphate charges corresponding to standard AMBER
values. The electrostatic interactions were treated by the SPME
method \cite{Essmann:95}, with 9 \AA\ truncation for the real space
sum and common values of other Ewald parameters \cite{Mzjacs:98}. The
temperature was maintained by the Berendsen algorithm
\cite{Berendsen:84} applied separately to solute and solvent with a
relaxation time of 10 ps.  The average temperature was about 298K. To
increase the time step, MD simulations were carried out by the
internal coordinate method (ICMD) \cite{Mzjcc:97,Mzjchp:99}, with the
internal DNA mobility limited to essential degrees of freedom. The
rotation of water molecules and internal DNA groups including only
hydrogen atoms was slowed down by weighting of the corresponding
inertia tensors \cite{Mzjacs:98,Mzjpc:98}. The double-helical DNA was
modeled with free backbone torsions and bond angles in sugar rings.
Phosphate groups and aromatic cycles were rigid. The net effect of
these constraints upon DNA dynamics is not significant, which was
checked earlier through comparisons with conventional Cartesian MD
\cite{Mzjacs:98,Mzbj:06}. The time step was 0.01 ps.  The statistics
of Poincar\'e recurrences $P(\tau)$ is a positive definite function
stable with respect to fluctuations in contrast to time
autocorrelation functions. Importantly, its numerical evaluation is
trivially parallelizable, that is, it can be accumulated in a large
number of independent MD trajectories. The results presented in the
text were obtained by using parallel computations on 129 cores for the
model system that had 3557 degrees of freedom for 1665 atoms to obtain
the total sampling of about 100 $\mu s$.

In the modeled DNA duplex, the base pairs flanking the coumarin probe
are next to the terminal ones that also can be opened. In such cases,
the base pair under study is effectively transformed into a terminal
one. In experiments, this duplex is part of a longer DNA and we would
like to simulate the corresponding environment. Moreover, for reliable
statistical averaging stable conditions are required, that is, the
dynamics should not involve slower motions that cannot be averaged
during in the same trajectory. The base flipping of terminal base
pairs can be discarded, for instance, by stopping the trajectory. A
more practical solution consists in applying flat-bottom restraints to
some hydrogen bonds in the external base pairs as explained elsewhere
\cite{Mzjctc:09}. The restraints are switched off when the distance
between the hydrogen bonded heavy atoms is below 3.8 \AA, that is,
they do not perturb DNA dynamics, but increase the probability of
closure when the hydrogen bond is opened. The internal base pairs were
free to open completely. In three such cases one of the residues in
the 3'n base pair flipped out and remained outside the stack. This did
not affect the results, which was checked by excluding these three
trajectories from the analysis. Other base pairs opened only
temporarily, with all such events reversed at the end.

\section{Statistics of Poincar\'e recurrences and autocorrelation
function}

In the chaos theory, the autocorrelation function
$C(\tau)$ and the probability $P(\tau)$ are related as
\cite{Karney:83,Chirikov:99}
\beq\label{ECtau1a}
C(\tau) \propto \tau P(\tau).
\eeq
This relationship follows from the following reasoning. Suppose we
have an indicator function $f(t)$ such that $f=1$ on the intervals
counted as Poincar\'e returns and $f=0$ elsewhere. The unnormalized
autocorrelation function is defined as
\beq\label{ECtau2}
C(\tau)=\int^{\infty}_{0}f(t)f(t+\tau)dt.
\eeq
Non-zero contributions to this integral come from returns longer
than $\tau$. Duration $t>\tau$ gives a contribution $t-\tau$ while
the number of such returns is $-dP(t)$, therefore,
\[
C(\tau)=-\int^{\infty}_{\tau}(t-\tau)P^{'}_{t}dt=
 -\int^{\infty}_{\tau}tP^{'}_{t}dt
 +\int^{\infty}_{\tau}\tau P^{'}_{t}dt.
\]
Assuming that $P(\tau)$ is smooth and integrable, the above integrals
readily result in
\beq\label{ECtau3}
C(\tau)=\int^{\infty}_{\tau}P(t)dt
\eeq
and \Req{Ctau1a} for the power-law decay.

The above derivation works for canonical chaotic systems
\cite{Karney:83,Chirikov:99} because trajectories are quasi-periodic
during long stays near stability islands and chaotic in between them.
Chaotic intervals do not contribute to \Req{Ctau2} and quasiperiodic
motion gives contributions proportional to the lengths of intervals.
Different quasi-periodic intervals are uncorrelated on average.  For
the distances measured in MD these conditions are violated already
because the averages grow with the duration of opening (see Fig. 3).
The power spectral density of time dependences decays with a power-law
exponent $\tau=1.63$ \cite{Mzprl:15}, which means that the
corresponding autocorrelation function cannot agree with \Req{Ctau1a}
under simple assumptions about the dimensionality of the space. This
problem can be obviated by assuming that the electrostatic energy term
that links the HB-breathing with the dye's dipole momentum is
proportional to an indicator function incremented by $\pm1$ when the
H-bond is broken and reformed, respectively. The corresponding
autocorrelation function should have a power-law tail with the desired
exponent. However, this is an {\em ad hoc} mathematical rewording of
the theory considered in the main text rather than a physical model
compatible with Eq. (3).  The agreement with experiment obtained in
this way has no physical meaning.

\section{Linear approximation and exponential relaxation}

In the linear case, the dynamics of the subset of excited
coumarins is described by schema
\beq
0\xleftarrow[]{k_0}
A\xrightleftharpoons[k_{-}]{k_{+}}B\xrightarrow[]{k_0}0,
\eeq
with the rate constants of base pair opening $k_{+}$ and $k_{-}$,
respectively. Constant $k_0$ describes quenching due to any process
and defines the excitation lifetime.  The corresponding exponential
decay is subtracted during the data processing and does not
contribute to the response function $S(t)$.  To discard it the
solution is written as ${\bf C}(t)exp(-k_0 t)$ where vector ${\bf
C}(t)$ corresponds to a simpler system
$A\xrightleftharpoons[k_{-}]{k_{+}}B$ described by equation
\beq\label{ECdot}
\dot{\bf C}={\bf M}{\bf C};\
{\bf C}=\begin{pmatrix*}C_A\\C_B\end{pmatrix*},
{\bf M}=\begin{pmatrix*}[r] -k_{+}&k_{-}\\k_{+}&-k_{-}\end{pmatrix*}
\eeq
where $C_A$ and $C_B$ are relative populations of states A and B,
respectively, while the overdot denotes time derivative.
The general solution of \Req{Cdot} is
\beq\label{EC=}
{\bf C}=C_1{\bf e}_1\exp\left(-\lambda_1 t\right)+C_2{\bf
e}_2\exp\left(-\lambda_2
t\right)
\eeq
with free constants $C_{1,2}$. The two eigenvalues and unnormalized
eigenvectors computed as
\beq
\lambda_1=0, {\bf e}_1=\begin{pmatrix*}k_{-}\\k_{+}\end{pmatrix*};\
\lambda_2=k_{+}+k_{-}, {\bf e}_2=\begin{pmatrix*}-1\\+1\end{pmatrix*}
\eeq
correspond to equilibrium and relaxation, respectively.
Substitution of \Req{C=} into Eq. (6) yields
\beq\label{ESt4}
S(t)=\exp\left(-\lambda_2 t\right).
\eeq

The exponential decay in \Req{St4} looks irreversible because vector
${\bf e}_2$ is antisymmetric, which corresponds to an excess in $B$
and a hole in $A$ or vice versa.  The hole is gradually closed without
creating an opposite flow.  If $k_{-}\gg k_{+}$ than
$\lambda_2=k_{-}$, that is, even if the perturbation is made by adding
molecules to $A$ the relaxation looks like a faster $B\rightarrow A$
transition. In this case the hole in $B$ is closed  by the equilibrium
$A\rightarrow B$ flow which is strong because $C^0_A\gg C^0_B$
according to vector ${\bf e}_1$.

With different emission spectra in states $A$ and $B$, a small
difference in the excitation lifetimes is sufficient to cause a TRFSS
response.  This mechanism is applicable even if the coumarin's dipole
moment is not changed by excitation. However, a large increase of the
dipole moment characteristic for coumarins can perturb the
$A\rightleftharpoons B$ equilibrium directly, which gives an
alternative physically distinct perturbation pathway. The mathematical
treatments of these two pathways are similar.

Suppose that the $S0\rightarrow S1$ excitation in coumarin shifts the
equilibrium, i.e., alters constants $k_{+}$ and/or
$k_{-}$. When the laser pulse arrives the system's eigenvalues and
eigenvectors are changed instantaneously, which starts the
relaxation to a new equilibrium described by an exponential decay with
new (primed) ${\bf e}'_1$ and $\lambda'_2$
\beq\label{EC=1}
{\bf C}=C^0{\bf e}'_1+\delta{\bf e}_2\exp\left(-\lambda'_2 t\right)
\eeq
where $C^0$ and $\delta$ the components of the initial state in
the primed basis. The same result is obtained if the excitation
lifetimes in states $A$ and $B$ are slightly different. Suppose the
quenching rates in states $A$ and $B$ are $k_0$ and
$k_0+\widetilde{k}$, respectively, with $\widetilde{k}\gtrapprox 0$.
Matrix $\bf M$ in \Req{Cdot} takes the form
\beq\label{EM=}
{\bf M}=\begin{pmatrix*}[r]
-k_{+}&k_{-}\\k_{+}&-(k_{-}+\widetilde{k})\end{pmatrix*}
\eeq
with slightly different eigenvalues and eigenvectors. As in the
previous case after the laser pulse the $A\xrightleftharpoons[]{}B$
dynamics in the subset of excited coumarins is out of equilibrium and
the relaxation occurs according to \Req{C=1}.

\bibliography{last}

\end{document}